\documentclass[conference]{IEEEtran}
\usepackage{amsmath,graphicx}
\usepackage{amssymb}
\usepackage{amsthm}
\usepackage{algorithm}
\usepackage{algorithmic}
\usepackage{array}
\usepackage{url}

\title{Graph-Based Radio Resource Management for Vehicular Networks}

\author{Le Liang$^\dag$, Shijie Xie$^{\ddag}$, Geoffrey Ye Li$^\dag$, Zhi Ding$^{\S}$, and Xingxing Yu$^{\ddag}$\\
$^\dag$ School of Electrical and Computer Engineering; $\ddag$ School of Mathematics \\
	Georgia Institute of Technology, Atlanta, GA\\
$^{\S}$ Department of Electrical and Computer Engineering,
	University of California, Davis, CA\\
    Email: \{lliang, shijie.xie\}@gatech.edu; liye@ece.gatech.edu;
   zding@ucdavis.edu; yu@math.gatech.edu
}

\begin{document}

\maketitle
\newtheorem{theorem}{Theorem}
\begin{abstract}
This paper investigates the resource allocation problem in device-to-device (D2D)-based vehicular communications, based on slow fading statistics of channel state information (CSI), to alleviate signaling overhead for reporting rapidly varying accurate CSI of mobile links.
We consider the case when each vehicle-to-infrastructure (V2I) link shares spectrum with multiple vehicle-to-vehicle (V2V) links.
Leveraging the slow fading statistical CSI of mobile links, we maximize the sum V2I capacity while guaranteeing the reliability of all V2V links.
We propose a graph-based algorithm that uses graph partitioning tools to divide highly interfering V2V links into different clusters before formulating the spectrum sharing problem as a weighted 3-dimensional matching problem, which is then solved through adapting a high-performance approximation algorithm.
\end{abstract}

\section{Introduction}
\label{sec:intro}
The emergence of vehicle-to-everything (V2X) communications aims to make everyday vehicular operation safer, greener, and more efficient, thus paving the path to autonomous driving with the advent of the fifth generation (5G) cellular system \cite{Araniti2013lte,Liang2017vehicular}.
Various communications standards, e.g., dedicated short range communications (DSRC) \cite{Kenney2011dedicated} and the intelligent transportation system (ITS)-G5 \cite{ITSG5}, both based on the IEEE 802.11p standard \cite{ieee2010ieee}, have been historically developed to ensure interoperability  in information exchange among vehicles.
However, recent studies \cite{Araniti2013lte,Sun2016support} have revealed several inherent issues of the 802.11p-based technology, including scalability, potentially unbounded channel access delay, lack of quality-of-service (QoS) guarantees, and short-lived vehicle-to-infrastructure (V2I) connection.
To address the issue, 3GPP has recently started working towards supporting vehicle-to-everything (V2X) services in long term evolution (LTE) networks \cite{Sun2016support,3GPPr14v2x}.
Widely deployed cellular networks, assisted with direct device-to-device (D2D) underlay communications \cite{Araniti2013lte,Feng2013device}, have shown significant potential in enabling efficient and reliable vehicle-to-vehicle (V2V) and V2I communications, meeting the diverse V2X QoS requirements and providing immunity to high mobility.

In this paper, we consider resource allocation for D2D-based vehicular networks, where each V2I link shares spectrum with multiple V2V links and different vehicular links have diverse QoS requirements.
We take advantage of both optimization and graph theoretic tools to develop a low-complexity algorithm to solve the problem.
In the proposed algorithm, we divide the V2V links into disjoint spectrum-sharing clusters using graph partitioning algorithms to mitigate their mutual interference.
We then model the spectrum allocation problem as a weighted 3-dimensional matching problem in graph theory, where weights of edges in the graph are obtained by optimizing powers of both V2I and V2V transmitters for each feasible spectrum sharing candidate.
Instead of using the local search based approximation \cite{berman2000d,Wei2017resource} for the proposed weighted 3-dimensional matching problem, we adopt the algorithm from \cite{chan2012linear}, which combines the iterative rounding and the fractional local ratio methods, and reduces the approximation factor from $(2+\epsilon)$ to $2$ in polynomial time with $ \epsilon > 0$.


\section{SYSTEM MODEL}\label{sec:sys}
\begin{figure}[!t]
\centering
\includegraphics[width=0.9\linewidth]{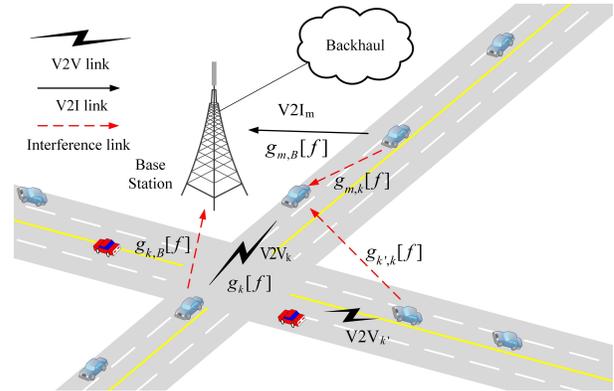}
\caption{D2D-based vehicular communications.}\label{fig:v2x}
\end{figure}

Consider a D2D-based vehicular communications network as shown in Fig.~\ref{fig:v2x}. There are $M$ V2I and $K$ V2V communication links.
The $M$ V2I links are initiated by $M$ single-antenna vehicles, demanding large-capacity uplink connection with the base station (BS) to support various bandwidth intensive applications..
The $K$ V2V links are formed among the vehicles, designed with high reliability such that safety critical information can be shared among neighboring vehicles reliably, in the form of localized D2D communications.

Denote the set of V2I links as ${\mathcal M} = \{1,\cdots,M\}$ and the set of V2V links as ${\mathcal K} = \{1,\cdots,K\}$.
The total available bandwidth is divided into $F$ resource blocks (RBs), denoted by $\mathcal{F} = \{1,\cdots,F\}$. Without loss of generality, we assume $M=F$ in this paper and each of the $M$ V2I links uses a single RB\footnote{We can define multiple virtual V2I links when the V2I link requires more than one RB.}, i.e., no spectrum sharing among V2I links.
To improve spectrum utilization, orthogonally allocated uplink spectrum of V2I links is reused by V2V links since uplink resource usage is less intensive and interference at the BS is more manageable.
We note that in practice the number of V2V links tends to be larger than that of V2I links, i.e., $K \gg M$, making spectrum reuse among V2V links necessary.

As in Fig.~\ref{fig:v2x}, the channel power gain, $g_{m,B}[f]$, from the transmitter of the $m$th V2I link to the BS over the $f$th RB is
\begin{align}\label{eq:channel}
g_{m,B}[f] = \alpha_{m,B}|h_{m,B}[f]|^2,
\end{align}
where $h_{m,B}[f]$ is the small-scale fading component, assumed to be distributed according to $\mathcal{CN}(0,1)$ and independent across different RBs and links, and $\alpha_{m,B}$ captures large-scale fading effects, i.e., including path loss and shadowing, assumed to be independent of the RB index $f$.
Similarly, we can define the $k$th V2V channel over the $f$th RB, $g_{k}[f]$, the interfering channel from the $k'$th V2V transmitter to the $k$th V2V receiver
over the $f$th RB, $g_{k',k}[f]$, the interfering channel from the $m$th V2I transmitter to the $k$th V2V receiver over the $f$th RB, $g_{m,k}[f]$, and the interfering channel from the $k$th V2V transmitter to the BS over the $f$th RB, $g_{k,B}[f]$.

The full CSI of links engaging the BS, including the V2I channels, $g_{m,B}[f]$, and the interfering channels from the V2V transmitters, $g_{k,B}[f]$, can be estimated at the BS, and is thus assumed known at the central controller.
However, the CSI of mobile links, including the V2V channels, $g_{k}[f]$, the peer V2V interfering channels, $g_{k',k}[f]$, and the interfering channels from the V2I transmitters, $g_{m,k}[f]$, has to be estimated at the mobile receiver and then reported to the BS periodically.
Frequent feedback of the fast fading information of rapidly varying mobile channels incurs substantial signaling overhead and thus makes tracking instantaneous CSI of mobile channels infeasible in practice. Therefore in this paper, we assume that the BS only has access to the large-scale fading information of such channels, which varies on a slow scale. In the meantime, each realization of the fast fading is unavailable at the BS while its statistical characterization is assumed to be known.

To this end, the received signal-to-interference-plus-noise ratios (SINRs) of the $m$th V2I link at the BS and the $k$th V2V link at the V2V receiver over the $f$th RB can be expressed as
\begin{align}
\gamma_{m,f}^c = \frac{P_{m,f}^cg_{m,B}[f]}{\sigma^2 + \sum\limits_{k}\rho_{k,f}^dP_{k,f}^d g_{k,B}[f]}
\end{align}
and
\begin{align}
\gamma_{k,f}^d = \frac{P_{k,f}^d g_{k}[f]}{\sigma^2 + \sum\limits_{m}\rho_{m,f}^c P_{m,f}^cg_{m,k}[f] + \sum\limits_{k'\ne k}\rho_{k',f}^d P_{k',f}^d g_{k',k}[f] },
\end{align}
respectively, where $P_{m,f}^c$ and $P_{k,f}^d$ denote transmit powers of the $m$th V2I transmitter and the $k$th V2V transmitter over the $f$th RB, respectively, $\sigma^2$ is the noise power, and $\rho_{m,f}^c\in \{0,1\}$ is the spectrum allocation indicator with $\rho_{m,f}^c=1$ implying the $m$th V2I links is transmitting over the $f$th RB and $\rho_{m,f}^c=0$ otherwise. The spectrum allocation indicator for the $k$th V2V link, $\rho_{k,f}^d$, is similarly defined.

To meet the diverse QoS requirements for different vehicular links, i.e., large capacity for V2I connections and high reliability for V2V connections, we maximize the sum capacity of the $M$ V2I links while guaranteeing the minimum reliability for each V2V link. The spectrum and power allocation problem is formulated as:

\begin{align}
 \max_{ \overset{\{\rho_{m,f}^c, \rho_{k,f}^d\}} {\{P_{m,f}^c,P_{k,f}^d\}} } & \sum\limits_{m}\sum\limits_{f}\rho_{m,f}^c \log_2(1 + \gamma_{m,f}^c ) \label{eqOpt}\\
 \text{s.t.} ~~~~ & \rho_{k,f}^d\text{Pr}\left\{\gamma_{k,f}^d \le \gamma_{0}^d \right\} \le p_0, \forall k,f \tag{\ref{eqOpt}a} \\
& \sum\limits_{m}\rho_{m,f}^c = 1, \forall f \tag{\ref{eqOpt}b} \\
& \sum\limits_{f}\rho_{m,f}^c = 1, \forall m \tag{\ref{eqOpt}c} \\
& \sum\limits_{f}\rho_{k,f}^d = 1, \forall k \tag{\ref{eqOpt}d} \\
& \sum\limits_{f}\rho_{m,f}^c P_{m,f}^c \le P_{\text{max}}^c, \forall m \tag{\ref{eqOpt}e} \\
& \sum\limits_{f} \rho_{k,f}^d P_{k,f}^d \le P_{\text{max}}^d, \forall k \tag{\ref{eqOpt}f} \\
& P_{m,f}^c \ge 0, P_{k,f}^d \ge 0, \forall m,k,f \tag{\ref{eqOpt}g} \\
& \rho_{m,f}^c, \rho_{k,f}^d \in \{0,1\}, \forall m,k,f \tag{\ref{eqOpt}h},
\end{align}
where $\gamma_0^d$ in (\ref{eqOpt}a) is the minimum SINR needed to establish a reliable V2V link and $p_0$ in (\ref{eqOpt}a) is the tolerable outage probability.
$P_{\text{max}}^c$ in (\ref{eqOpt}e) and $P_{\text{max}}^d$ in (\ref{eqOpt}f) are the maximum transmit powers of the V2I and V2V transmitters, respectively.
Constraint (\ref{eqOpt}a) represents the minimum reliability requirement for $K$ V2V links, where the probability is evaluated in terms of the random fast fading of mobile channels.
Constraint (\ref{eqOpt}b) restricts orthogonal spectrum to be allocated among $M$ V2I links.
Constraints (\ref{eqOpt}c) and (\ref{eqOpt}d) model our assumption that each of the V2I and V2V links accesses a single RB.
Constraints (\ref{eqOpt}e) and (\ref{eqOpt}f) ensure the transmit powers of V2I and V2V links cannot go beyond their maximum limits.

\section{SPECTRUM AND POWER ALLOCATION}\label{sec:allocation}

The optimization problem in \eqref{eqOpt} is combinatorial in nature and is further complicated by the nonlinear constraints and objective function. To address the problem, we first exploit graph partitioning algorithms to divide the V2V links into different clusters based on their mutual interference. This will identify proper V2V sets for spectrum sharing with minimum interference.
Next, all V2V links in each cluster are allowed to share the same spectrum with one of the $M$ V2I links while V2V links in different clusters cannot share spectrum. We then optimize V2I and V2V transmit powers for all possible sharing patterns.
Finally, we construct a 3-partite graph, with the $M$ V2I links, $F$ RBs, and $N$ V2V clusters as its vertices and with edge weights equal to the V2I capacity from applying optimized V2I and V2V transmit powers. The resource allocation problem in \eqref{eqOpt} can then be reduced to a weighted 3-dimensional matching problem.

\subsubsection{V2V Partitioning}
\begin{figure}
\centering
\includegraphics[width=0.5\linewidth]{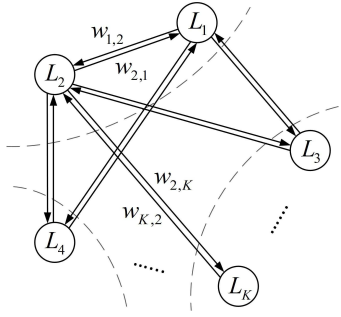}
\caption{Graph representation for interfering links.}\label{fig:graphLinks}
\end{figure}

The interference management for V2V links can be captured using a graph as in Fig.~\ref{fig:graphLinks}, where each V2V link $L_k$ is modeled as a vertex and two vertices are joined by an edge when they are mutually interfering. The edge weight is set to capture the interference level with $w_{k',k}=\alpha_{k',k}$, where $\alpha_{k',k}$ is the large-scale fading CSI of the interference channel from the $k'$th V2V transmitter to the $k$th V2V receiver.
The goal is to partition the $K$ vertices into $N$ sets, $C_1, \cdots, C_N$, where $N\ll K$, minimizing the intra-cluster interference across all clusters, i.e., $\sum\limits_n\left(\sum\limits_{k',k\in C_n}w_{k',k}\right)$.
Intuitively, this implies that we attempt to partition strongly interfering V2V links into different sets so that links within the same set can share the same RB without incurring too much mutual interference.

The above partitioning problem is equivalent to the MAX $N$-CUT problem in graph theory \cite{Sahni1976Pcomplete,Chang2009multicell} and a brief explanation is given here.
Let $G$ be a graph with vertex set $V(G)$ and edge set $E(G)$. Let $w: E(G) \rightarrow \mathbb{R}$.
The MAX $N$-CUT problem for a weighted graph is to find a partition of the graph $G$ into $N$ disjoint clusters $C_n, n = 1, \cdots,N$, such that $C_1 \cup \cdots \cup C_N = V(G)$ and $\sum\limits_{a\in C_i, b\in C_j, i<j}w_{a,b}$ is maximized, where $w_{a,b}$ is the weight of the edge $(a,b)$.
Since $\sum\limits_n\left(\sum\limits_{k',k\in C_n}w_{k',k}\right) + \sum\limits_{a\in C_i, b\in C_j, i<j}w_{a,b} = \sum\limits_{e\in E(G)} w(e)$, maximizing $\sum\limits_{a\in C_i, b\in C_j, i<j}w_{a,b}$ is thus equivalent to minimizing the other term.

A simple heuristic algorithm has been proposed in \cite{Sahni1976Pcomplete} and exploited for interference management in \cite{Chang2009multicell} for multicell OFDMA systems, achieving an absolute ratio of $(1-1/N)$ for a general $N$-CUT problem. This algorithm is listed in Table~\ref{alg:maxCut} and will be used in this paper.

\begin{table}[!t]
\caption{Heuristic Algorithm for MAX $N$-CUT \cite{Chang2009multicell,Sahni1976Pcomplete}} \label{alg:maxCut} \centering
\begin{algorithm}[H]
\caption{Heuristic Algorithm for V2V Partitioning} \label{algm:maxCut}
\begin{algorithmic}[1]
\normalsize{
\STATE Arbitrarily assign one V2V link to each of the $N$ clusters.
\FOR{$k \in \mathcal{K}$ \textbf{and} not already in any cluster}
\FOR{$n=1:N$}
\STATE Compute the increased intra-cluster interference using $\sum\limits_{k'\in C_n} (w_{k,k'} + w_{k',k})$.
\ENDFOR
\STATE Assign the $k$th V2V link to the $n^*$th cluster with $n^* = \text{arg} \min \sum\limits_{k'\in C_n} (w_{k,k'} + w_{k',k})$.
\ENDFOR
\STATE Return the V2V clustering result.
}
\end{algorithmic}
\end{algorithm}
\end{table}

\subsubsection{Power Allocation Design}\label{sec:power}
As mentioned before, V2V links in one cluster can share the spectrum with one V2I link while those in different clusters cannot share. For an arbitrary spectrum sharing pattern, e.g., when the $m$th V2I link is transmitting over the $f$th RB, which is shared by all V2V links in the $n$th cluster, we attempt to find its optimal power control for both V2I and V2V links. That is, we maximize the V2I capacity, defined as $R_{m,n}[f]$, with the reliability of all V2V links in the $n$th cluster guaranteed when they share the $f$th RB. The power optimization problem is formulated as
\begin{align}
 \max\limits_{P_{m,f}^c,\{P_{k,f}^d\}} & \log_2\Bigg(1 + \frac{P_{m,f}^c g_{m,B}[f]}{\sigma^2 + \sum\limits_{k\in C_n}P_{k,f}^d g_{k,B}[f]} \Bigg) \triangleq R_{m,n}[f] \label{eq1on1} \\
 \text{s.t.} ~~~~~ & \text{Pr}\left\{\frac{P_{k,f}^d g_{k}[f]}{\sigma^2 + P_{m,f}^c g_{m,k}[f] + \sum\limits_{k'\ne k}P_{k',f}^d g_{k',k}[f] } \le \gamma_{0}^d \right\} \nonumber\\
 & \hspace{4.1cm} \le p_0, \forall k \in C_n \tag{\ref{eq1on1}a}  \\
& 0 \le P_{m,f}^c \le P^c_{\text{max}} \tag{\ref{eq1on1}b}\\
& 0 \le P_{k,f}^d \le P^d_{\text{max}}, \forall k \in C_n \tag{\ref{eq1on1}c}.
\end{align}

The reliability constraint (\ref{eq1on1}a) can be manipulated and turned into an analytic (tight) upper bound based on the results developed in \cite{Kandukuri2002optimal,Papandriopoulos2005optimal}. Due to page limit, we omit the details and directly present the transformed optimization problem as
\begin{align}
\max\limits_{P_{m,f}^c,\{P_{k,f}^d\}} & \log_2\Bigg(1 + \frac{P_{m,f}^c g_{m,B}[f]}{\sigma^2 + \sum\limits_{k\in C_n}P_{k,f}^d g_{k,B}[f]} \Bigg)  \label{eq1on1Upper} \\
\text{s.t.} ~~~~~ &  \frac{P_{k,f}^d\alpha_{k}}{\sigma^2 + P_{m,f}^c \alpha_{m,k} +\sum\limits_{k'\ne k} P_{k',f}^d\alpha_{k',k}} \ge \frac{\gamma_0^d}{\ln\frac{1}{1-p_0}} , \nonumber\\
& \hspace{4.5cm} \forall k \in C_n \tag{\ref{eq1on1Upper}a}  \\
& 0 \le P_{m,f}^c \le P^c_{\text{max}} \tag{\ref{eq1on1Upper}b}\\
& 0 \le P_{k,f}^d \le P^d_{\text{max}}, \forall k \in C_n. \tag{\ref{eq1on1Upper}c}
\end{align}

We derive the optimal solution to the above optimization problem in the following theorem with detailed proof omitted.
\begin{theorem}
The optimal solution to \eqref{eq1on1Upper} is given by\footnote{Should either the optimal V2I transmit power, i.e., $P_{m,f}^{c^*}$, or any of the optimal V2V transmit powers in the $n$th cluster, i.e., ${P}_{k,f}^{d^*}, k\in C_n$, be negative, we declare the problem in \eqref{eq1on1Upper} infeasible and set $R_{m,n}[f] = -\infty$.}
\begin{align}\label{eq:optPc}
P_{m,f}^{c^*} = \min \left\{P_{\text{max}}^c, \left\{ \frac{P^d_{\text{max}} - \bar{\gamma_0}\sigma^2\pmb{\phi}_i^H \mathbf {1}}{\bar{\gamma_0}\pmb{\phi}_i^H\pmb{\alpha}_m} \right\}_{i=1}^{N_{c_n}} \right\},
\end{align}
and
\begin{align}\label{eq:optPd}
\mathbf{P}_{n,f}^{d^*} = \mathbf{\Phi}^{-1} \bar{\gamma}_0 \left( P_{m,f}^{c^*} \pmb{\alpha}_m  + \sigma^2  \right),
\end{align}
where $\mathbf{P}_{n,f}^d\in\mathbb{R}^{N_{c_n}\times 1}$ stores transmit powers of all $N_{c_n}$ V2V links in the $n$th cluster, $\bar{\gamma}_0 = \frac{\gamma_0^d}{-\ln(1-p_0)}$, $\pmb{\alpha}_m = (\alpha_{m,1}, \cdots, \alpha_{m,N_{c_n}})^T \in \mathbb{C}^{N_{c_n}\times 1}$, $\mathbf{1}$ is an all-one vector, $\mathbf{\Phi}\in\mathbb{C}^{N_{c_n}\times N_{c_n}}$ is given by
\begin{align}
\mathbf{\Phi}_{i,j} =
\begin{cases}
\alpha_i, & \text{if $i=j$},\\
-\bar{\gamma}_0\alpha_{j,i}, & \text{otherwise},
\end{cases}
\end{align}
and $\pmb{\phi}_i^H$ is the $i$th row of the inverse of $\pmb{\Phi}$, i.e., $\pmb{\Phi}^{-1}$.
\end{theorem}

\subsubsection{Resource Matching} \label{sec:resourceMatching}

\begin{figure}
\centering
\includegraphics[width=0.7\linewidth]{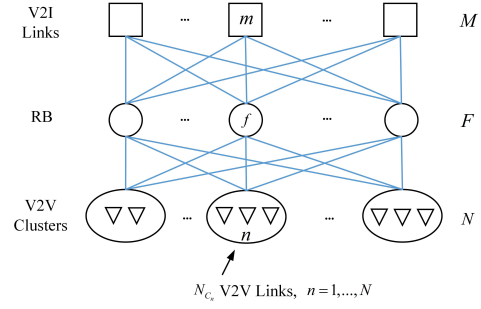}
\caption{Graph representation for spectrum sharing among V2I and V2V links.}\label{fig:matching}
\end{figure}

To this end, essential elements of the resource allocation problem in \eqref{eqOpt} can be modeled as a 3-partite graph in Fig.~\ref{fig:matching}. For each of the possible V2I-RB-V2V resource sharing patterns ($MFN$ in total), we formulate the optimization problem as in \eqref{eq1on1} and then find the resulting V2I capacity $R_{m,n}[f], \forall m,n,f$. The weight for the edge linking from the $m$th V2I vertex in the upper layer, through the $f$th RB vertex in the middle layer, and to the $n$th cluster vertex in the lower layer, is set to be $R_{m,n}[f]$. Then the spectrum allocation problem reduces to
\begin{align}
\max\limits_{\{\rho_{m,f}^c,\rho_{n,f}^{cl}\}} &  \sum\limits_{m}\sum\limits_{f}\sum\limits_{n}\rho_{m,f}^c\rho_{n,f}^{cl} R_{m,n}[f] \label{eqMatch}\\
\text{s.t.} ~~~~ &  \sum\limits_{m}\rho_{m,f}^c = 1,  \sum\limits_{n}\rho_{n,f}^{cl} = 1, \forall f \tag{\ref{eqMatch}a} \\
& \sum\limits_{f}\rho_{m,f}^c = 1, \forall m, \sum\limits_{f}\rho_{n,f}^{cl} = 1, \forall n \tag{\ref{eqMatch}b} \\
& \rho_{m,f}^c,\rho_{n,f}^{cl} \in\{0,1\}, \forall m,n,f  \tag{\ref{eqMatch}c}
\end{align}

This problem can be transformed into a weighted 3-dimensional matching problem with weights of $w(m,f, n)=R_{m, n}[f]$, for $1\leq m\leq M,1\leq f \leq F,$ and $1\leq n\leq N$, with a brief explanation given as follows.
We first construct a $3$-partite hypergraph $H=(V, E)$, by letting $V=\{[m,0,0]: 1\leq m\leq M\}\cup \{[0,f,0]: 1\leq f\leq F\}\cup\{[0,0,n]: 1\leq n\leq N\}$, and $E=\{(m, f, n): 1\leq m\leq M, 1\leq f\leq F, 1\leq n\leq N\}$, where  $(m,f,n)=\{[m, 0, 0], [0,f,0], [0,0,n]\}$.
We define the weight function $w: E\rightarrow \mathbb{R}$ by letting $w(m,f, n)=R_{m, n}[f]$, for all $1\leq m\leq M,1\leq f \leq F,  1\leq n\leq N.$ Now, we can see that solving our V2I-RB-V2V resource allocation problem is equivalent to solving the weighted $3$-dimensional matching problem on $H=(V, E)$ with weight function $w$.

Let $H=(V, E)$ be a 3-partite hypergraph, and let $w: E\rightarrow \mathbb{R}$. For $v\in V$, let $\delta(v)$ be the set of edges containing $v$. The weighted 3-dimensional matching problem can be formulated as the following integer program:
\begin{displaymath}
\begin{split}
\text{max }& \sum_{e\in E}w(e)x(e)\\
\text{s.t. }& \sum_{e\in \delta(v)}x(e)\leq 1, \forall v\in V\\
& x(e)\in \{0,1\}, \forall e\in E.
 \end{split}
\end{displaymath}
The linear programming relaxation of this integer program is
\begin{align}
\text{max }& \sum_{e\in E}w(e)x(e) \label{eq:relaxation} \\
\text{s.t. }& \sum_{e\in \delta(v)}x(e)\leq 1, \forall v\in V \nonumber\\
& x(e)\geq 0, \forall e\in E. \nonumber
\end{align}

To this end, we introduce the weighted 3-dimensional matching algorithm in Table~\ref{alg:3dMatching}.
Algorithm 2 in Table~\ref{alg:3dMatching} is obtained from the weighted 3-dimensional matching algorithm from \cite{chan2012linear} by adding Step 10. For any $e\in E$, let $N[e]$ be the set of edges of $H$ having nonempty intersection with $e$. Note that $e\in N[e]$.
In Algorithm~\ref{algm:3dMatching}, the solution $x$ of linear program~\eqref{eq:relaxation} must be basic; or else in Step 4, one cannot guarantee the existence of an edge $e\in E-F$ such that $x(N[e]\cap (E-F))\leq 2$. To obtain a basic solution of the linear program \eqref{eq:relaxation}, we could use some existing linear programming algorithm, such as the simplex algorithm or the dual-simplex algorithm. Indeed, any linear programming algorithm, which produces a basic solution, can be used here. We modified the algorithm in \cite{chan2012linear} by adding Step 10, because the original algorithm does not necessarily produce a maximal matching in $H$, since it only guarantees a matching with weights at least one half of the optimum. So, here, we use greedy algorithm to test whether or not $M_0$ is a maximal matching. If not, then we will find $E'$, with $w(e)\ge 0$ $\forall e\in E'$, such that $M_0\cup E'$ is a matching. In some cases, this added step could greatly improve the performance of the whole algorithm.

\begin{table}[!t]
\caption{Weighted 3-Dimensional Matching Algorithm \cite{chan2012linear}} \label{alg:3dMatching} \centering
\begin{algorithm}[H]
\caption{Weighted 3-Dimensional Matching Algorithm}\label{algm:3dMatching}
\begin{algorithmic}[1]
\normalsize{
\STATE Input: $H=(V, E), w:E\rightarrow \mathbb{R}$ and $x$, where $x$ is a basic solution of linear program \eqref{eq:relaxation} obtained by some linear programming algorithm.
\STATE Let $F \subseteq E$ with initialization $F=\emptyset$.

\REPEAT
\STATE Search for an edge $e\in E-F$ such that $x(N[e]\cap (E-F))\leq 2$.
\STATE Let $F=F\cup\{e\}$.
\STATE Let $i=|F|+1$, and let $i$ be the index of $e$.
\UNTIL{$E-F=\emptyset$}

\STATE Implement Local-Ratio algorithm in Table~\ref{alg:local} with input $F$ and $w$, where $w$ is the weight function on the edges of $H$.

\STATE Let $M_0$ be the output of Local-Ratio algorithm.

\STATE Use the greedy algorithm to find a maximal set $E'$ of edges, such that $M_0\cup E'$ is a matching, and $w(e)\ge 0$  for all $e\in E'.$ Then let $M_0\leftarrow M_0\cup E'$, and output $M_0$.
}
\end{algorithmic}
\end{algorithm}
\end{table}

\begin{table}[!t]
\caption{Local Ratio Algorithm \cite{chan2012linear}} \label{alg:local} \centering
\begin{algorithm}[H]
\caption{Local Ratio Algorithm \cite{chan2012linear}}
\begin{algorithmic}[1]
\normalsize{
\STATE Input: Hypergraph $H=(V, E)$, $F\subseteq E$, $w: E\rightarrow \mathbb{R}$, and an ordering of the edges in $E$.
\STATE Let $F'=\{e\in F: w(e)>0\}$.

\IF{ $F'=\emptyset$}
\STATE Return $\emptyset$.
\ENDIF

\STATE Let $e'$ be the smallest edge in $F'$ based on the ordering of $E$. Decompose the weight function $w=w_1+w_2$, where
$$w_1(e)=\left\{
    \begin{array}{c}
      w(e'), \text{ if } e\in N[e'].   \\
      0, \text{ otherwise.} \\
    \end{array}
    \right.$$

\STATE $M'\leftarrow$Local-Ratio($F', w_2$). (Note: this is a recursion.)

\IF{$M'\cup \{e'\}$ is a matching in $H$}
\STATE Return $M'\cup \{e'\}$.
\ELSE
\STATE Return $M'$.
\ENDIF
}
\end{algorithmic}
\end{algorithm}
\end{table}

Finally, the proposed graph-based algorithm to solve the problem in \eqref{eqOpt} is summarized in Table~\ref{tab:baseline}.

\begin{table}[!t]
\caption{Baseline Graph-based Resource Allocation} \label{tab:baseline} \centering
\begin{algorithm}[H]
\caption{Baseline Graph-based Resource Allocation} \label{algm:baseline}
\begin{algorithmic}[1]
\normalsize{
\STATE Use Algorithm~\ref{algm:maxCut} to divide $K$ V2V links into $N$ clusters, denoted by $C_1,\cdots, C_N$.
\FOR{ $m = 1:M$}
\FOR{ $n = 1:N$}
\FOR{ $f = 1:F$}
\STATE Use \eqref{eq:optPc} and \eqref{eq:optPd} to find the optimal V2I and V2V transmit powers, respectively.
\STATE Compute the V2I capacity, $R_{m,n}[f]$, with the optimized power control parameters.
\ENDFOR
\ENDFOR
\ENDFOR
\STATE Construct a 3-partite graph, where the $M$ V2I links, $F$ RBs, and $N$ V2V clusters form the vertices in three layers and the weight for each V2I-RB-V2V edge is set to $R_{m,n}[f]$.
\STATE Use Algorithm~\ref{algm:3dMatching} to find a matching solution $M_0$.
\STATE Return the 3-dimensional matching (spectrum sharing) result $M_0$ and the corresponding power allocation $\{(P_{m,f}^{c^*}, P_{k,f}^{d^*})\}$.
}
\end{algorithmic}
\end{algorithm}
\end{table}

\section{SIMULATION RESULTS}\label{sec:simulation}

\begin{table}[!t]
\scriptsize
\centering
\caption{Simulation Parameters \cite{3GPPr14v2x,3GPPsimulation}}\label{table:simulaton}
\begin{tabular}{|m{0.5\linewidth}|m{0.3\linewidth}|}
\hline
{\textbf{Parameter}} & \textbf{Value} \\\hline
Carrier frequency & 2 GHz \\\hline
Bandwidth & 10 MHz \\\hline
Cell radius  & 500 m      \\\hline
BS antenna height & 25 m \\\hline
BS antenna gain & 8 dBi \\ \hline
BS receiver noise figure & 5 dB \\ \hline
Distance from BS to highway & 35 m \\\hline
Vehicle antenna height & 1.5 m \\\hline
Vehicle antenna gain & 3 dBi \\\hline
Vehicle receiver noise figure & 9 dB \\\hline
Absolute vehicle speed $v$ & 70 km/h \\\hline
Vehicle drop model & spatial Poisson process \\ \hline
Number of lanes & 3 in each direction (6 in total) \\ \hline
Lane width & 4 m \\ \hline
Average inter-vehicle distance & $2.5v$, $v$ in m/s. \\ \hline
SINR threshold for V2V $\gamma_0^d$ & 5 dB \\ \hline
Reliability for V2V $p_0$ & 0.01 \\ \hline
Number of V2I links $M$ & 10\\ \hline
Number of V2V links $K$ & 30\\ \hline
Maximum V2I transmit power $P_{\text{max}}^c$ & 23 dBm \\ \hline
Maximum V2V transmit power $P_{\text{max}}^d$ & 23 dBm \\ \hline
Noise power $\sigma^2$  & -114 dBm \\ \hline
\end{tabular}
\end{table}

\begin{table}[!t]
\scriptsize
\centering
\caption{Channel Models for V2I and V2V Links \cite{3GPPr14v2x}}\label{table:channel}
\begin{tabular}{|m{0.26\linewidth}|m{0.28\linewidth}|m{0.24\linewidth}|}
\hline
\textbf{Parameter} & \textbf{V2I Link} & \textbf{V2V Link} \\ \hline
Pathloss model & 128.1 + 37.6$\log_{10}d$, $d$ in km  & LOS in WINNER + B1 \cite{WINNER}\\ \hline
Shadowing distribution & Log-normal & Log-normal\\ \hline
Shadowing standard deviation $\xi$ & 8 dB & 3 dB \\ \hline
Fast fading & Rayleigh fading & Rayleigh fading \\ \hline
\end{tabular}
\end{table}

\normalsize

In this section, simulation results are presented to validate the proposed resource allocation algorithm for D2D-based vehicular networks.
We follow the simulation setup for freeway detailed in 3GPP TR~36.885 \mbox{\cite{3GPPr14v2x}} and model a multi-lane freeway that passes through a single cell where the BS is located at its center as illustrated in Fig.~\ref{fig:v2x}.
The vehicles are dropped on the roads according to spatial Poisson process and the vehicle density is determined by the vehicle speed.
The $M$ V2I links are randomly chosen among generated vehicles and the $K$ V2V links are formed between each of the V2I links with its closest surrounding neighbors.
The major simulation parameters are listed in Table~\ref{table:simulaton} and the channel models for V2I and V2V links are described in Table~\ref{table:channel}. Note that all parameters are set to the values specified in Tables~\ref{table:simulaton} and \ref{table:channel} by default. In the simulation, the number of V2V clusters, $N$, is set to be equal to the number of V2I links, $M$.

\begin{figure}[!t]
\centering
\includegraphics[width=0.7\linewidth]{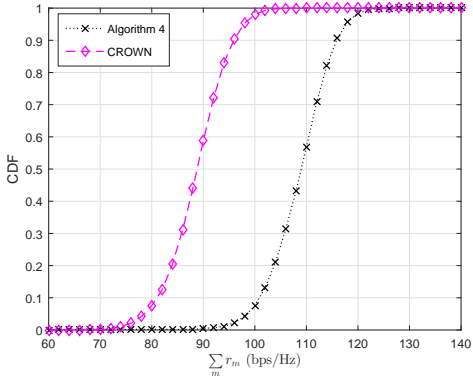}
\caption{CDF of instantaneous sum V2I capacity with Rayleigh fading.}\label{figCDFvsRate}
\end{figure}

Fig.~\ref{figCDFvsRate} compares the cumulative distribution function (CDF) of the instantaneous sum V2I capacity achieved by the proposed graph-based algorithm against the benchmark CROWN scheme developed in \cite{Sun2016cluster}.
We observe that the proposed Algorithm~4 outperforms the benchmark CROWN scheme since Algorithm~4 uses the slow fading CSI of mobile links while adapting to the fast fading CSI of links involving the BS.
In contrast, the benchmark CROWN scheme only adapts to the slow fading CSI of all links in the system.

\begin{figure}
\centering
\includegraphics[width=0.7\linewidth]{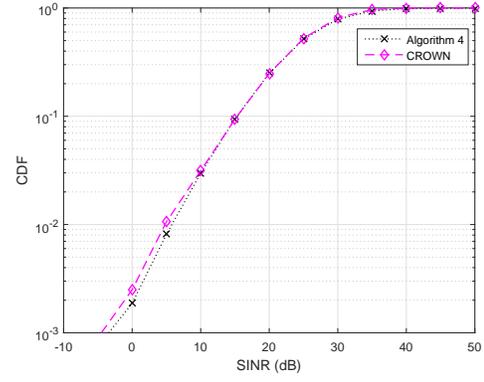}
\caption{CDF of instantaneous SINR of V2V links with Rayleigh fading, $\gamma_0^d = 5$ dB, and $p_0 = 0.01$.}\label{figCDFvsSINR}
\end{figure}

The reliability of V2V links is demonstrated in Fig.~\ref{figCDFvsSINR}, where the CDF of the instantaneous SINR of an arbitrary V2V link has been plotted.
From the figure, the proposed algorithm and the benchmark CROWN scheme achieve the SINR threshold, $\gamma_0^d = 5$ dB, at the targeted outage probability of $p_0=0.01$, justifying the effectiveness of the reliability guarantee of the proposed resource allocation scheme.

\section{CONCLUSIONS}\label{sec:conclusion}
This paper studied the resource allocation problem in D2D-based vehicular networks where each V2I link shares spectrum with multiple V2V links and the BS only has access to the slow fading CSI of vehicular links except those terminating at the BS.
We use graph partitioning tools to divide V2V links into disjoint clusters to minimize interference before formulating the spectrum allocation problem as a weighted 3-dimensional matching problem, tackled through adapting a high performance approximation algorithm.


\bibliographystyle{IEEEtran}
\bibliography{Ref}

\begin{thebibliography}{10}
\providecommand{\url}[1]{#1}
\csname url@samestyle\endcsname
\providecommand{\newblock}{\relax}
\providecommand{\bibinfo}[2]{#2}
\providecommand{\BIBentrySTDinterwordspacing}{\spaceskip=0pt\relax}
\providecommand{\BIBentryALTinterwordstretchfactor}{4}
\providecommand{\BIBentryALTinterwordspacing}{\spaceskip=\fontdimen2\font plus
\BIBentryALTinterwordstretchfactor\fontdimen3\font minus
  \fontdimen4\font\relax}
\providecommand{\BIBforeignlanguage}[2]{{%
\expandafter\ifx\csname l@#1\endcsname\relax
\typeout{** WARNING: IEEEtran.bst: No hyphenation pattern has been}%
\typeout{** loaded for the language `#1'. Using the pattern for}%
\typeout{** the default language instead.}%
\else
\language=\csname l@#1\endcsname
\fi
#2}}
\providecommand{\BIBdecl}{\relax}
\BIBdecl

\bibitem{Araniti2013lte}
G.~Araniti, C.~Campolo, M.~Condoluci, A.~Iera, and A.~Molinaro, ``{LTE for
  vehicular networking: A survey},'' \emph{IEEE Commun. Mag}, vol.~51, no.~5,
  pp. 148--157, May 2013.

\bibitem{Liang2017vehicular}
L.~Liang, H.~Peng, G.~Y. Li, and X.~Shen, ``Vehicular communications: A
  physical layer perspective,'' \emph{\emph{to appear in} IEEE Trans. Veh.
  Technol.}, 2017.

\bibitem{Kenney2011dedicated}
J.~B. Kenney, ``{Dedicated short-range communications (DSRC) standards in the
  United States},'' \emph{Proc. IEEE}, vol.~99, no.~7, pp. 1162--1182, Jul.
  2011.

\bibitem{ITSG5}
\emph{Intelligent Transport Systems (ITS); Cooperative ITS (C-ITS); Release 1},
  ETSI TR 101 607 V1.1.1, May 2013. [Online]. Available:
  \url{http://www.etsi.org/deliver/etsi_tr/101600_101699/101607/01.01.01_60/tr_101607v010101p.pdf}.

\bibitem{ieee2010ieee}
\emph{{IEEE Standard for Information Technology--Telecommunications and
  information exchange between systems--Local and metropolitan area
  networks--Specific requirements--Part 11: Wireless LAN Medium Access Control
  (MAC) and Physical Layer (PHY) specifications Amendment 6: Wireless Access in
  Vehicular Environments}}, IEEE Std. 802.11p-2010, Jul. 2010.

\bibitem{Sun2016support}
S.-H. Sun, J.-L. Hu, Y.~Peng, X.-M. Pan, L.~Zhao, and J.-Y. Fang, ``{Support
  for vehicle-to-everything services based on LTE},'' \emph{IEEE Wireless
  Commun.}, vol.~23, no.~3, pp. 4--8, Jun. 2016.

\bibitem{3GPPr14v2x}
\emph{{3rd Generation Partnership Project; Technical Specification Group Radio
  Access Network; Study on LTE-based V2X Services; (Release 14)}}, 3GPP TR
  36.885 V2.0.0, Jun. 2016.

\bibitem{Feng2013device}
D.~Feng, L.~Lu, Y.~Yuan-Wu, G.~Y. Li, G.~Feng, and S.~Li, ``{Device-to-device
  communications underlaying cellular networks},'' \emph{IEEE Trans. Commun.},
  vol.~61, no.~8, pp. 3541--3551, Aug. 2013.

\bibitem{berman2000d}
P.~Berman, ``A d/2 approximation for maximum weight independent set in d-claw
  free graphs,'' \emph{Algorithm Theory-SWAT 2000}, pp. 31--40, Jul. 2000.

\bibitem{Wei2017resource}
Q.~Wei, W.~Sun, B.~Bai, L.~Wang, E.~G. Str{\"{o}}m, and M.~Song, ``{Resource
  allocation for V2X communiations: A local search based 3D matching
  approach},'' in \emph{Proc. ICC}, May 2017, pp. 1--6.

\bibitem{chan2012linear}
Y.~H. Chan and L.~C. Lau, ``On linear and semidefinite programming relaxations
  for hypergraph matching,'' \emph{Mathematical Programming}, vol. 135, no.
  1-2, pp. 123--148, Oct. 2012.

\bibitem{Sahni1976Pcomplete}
S.~Sahni and T.~Gonzalez, ``P-complete approximation problems,'' \emph{J.
  Assoc. Comput. Mach.}, vol.~23, no.~3, pp. 555--565, Jul. 1976.

\bibitem{Chang2009multicell}
R.~Y. Chang, Z.~Tao, J.~Zhang, and C.-C.~J. Kuo, ``{Multicell OFDMA downlink
  resource allocation using a graphic framework},'' \emph{IEEE Trans. Veh.
  Technol.}, vol.~58, no.~7, pp. 3494--3507, Sep. 2009.

\bibitem{Kandukuri2002optimal}
S.~Kandukuri and S.~Boyd, ``{Optimal power control in interference-limited
  fading wireless channels with outage-probability specifications},''
  \emph{IEEE Trans. Wireless Commun.}, vol.~1, no.~1, pp. 46--55, Jan. 2002.

\bibitem{Papandriopoulos2005optimal}
J.~Papandriopoulos, J.~Evans, and S.~Dey, ``{Optimal power control for
  Rayleigh-faded multiuser systems with outage constraints},'' \emph{IEEE
  Trans. Wireless Commun.}, vol.~4, no.~6, pp. 2705--2715, Nov. 2005.

\bibitem{3GPPsimulation}
R1-165704, \emph{{WF on SLS evaluation assumptions for eV2X}}, 3GPP TSG RAN WG1
  Meeting \#85, May 2016.

\bibitem{WINNER}
\emph{{WINNER II Channel Models}}, IST-4-027756 WINNER II D1.1.2 V1.2, Sep.
  2007. [Online]. Available:
  \url{http://projects.celtic-initiative.org/winner+/WINNER2-Deliverables/D1.1.2v1.1.pdf}.

\bibitem{Sun2016cluster}
W.~Sun, D.~Yuan, E.~G. Str{\"o}m, and F.~Br{\"a}nnstr{\"o}m, ``Cluster-based
  radio resource management for {D2D}-supported safety-critical {V2X}
  communications,'' \emph{IEEE Trans. Wireless Commun.}, vol.~15, no.~4, pp.
  2756--2769, Apr. 2016.

\end{thebibliography}

\end{document}